\documentstyle[prl,aps]{revtex}

\newcommand{\gtilde}
 {~ \raisebox{-1ex}{$\stackrel{\textstyle >}{\sim}$} ~}
\newcommand{\ltilde}
 {~ \raisebox{-1ex}{$\stackrel{\textstyle <}{\sim}$} ~}

\begin{document}

\title{Gamma-Ray Bursts, Ultra High Energy Cosmic Rays, and Cosmic
Gamma-Ray Background} 

\author{Tomonori Totani}
\address{Department of Physics, School of Science, The University of Tokyo \\
Tokyo 113-0033, Japan \\
e-mail: totani@utaphp2.phys.s.u-tokyo.ac.jp}

\date{\today}

\maketitle

\begin{abstract}
We argue that gamma-ray bursts (GRBs) may be the origin of the
cosmic gamma-ray background radiation observed in GeV range.
It has theoretically been discussed that 
protons may carry a much larger amount of energy
than electrons in GRBs, and this large energy can be
radiated in TeV range by synchrotron radiation of ultra-high-energy
protons ($\sim 10^{20}$ eV). The possible detection of GRBs above 10 TeV
suggested by the Tibet and HEGRA groups also supports this idea.
If this is the case, most of TeV gamma-rays from GRBs are absorbed 
in intergalactic fields and eventually form GeV gamma-ray background,
whose flux is in good agreement with the recent observation.
\end{abstract}



\section{Introduction}
In all wavelengths from radio to gamma-rays, the presence of a diffuse
background radiation emanating from beyond our Galaxy is known. 
It is considered that
the background radiation in different bands has different origins, and
they may be truly diffuse processes or be superpositions of unresolved 
point sources. The background radiation in the highest energy range 
currently observed starts at around 30 MeV, and extends beyond 10 GeV
with a clear single power-law spectrum. This extragalactic background radiation
in the GeV range was first indicated by the SAS 2 satellite after removing the
much stronger GeV background from our Galaxy \cite{SAS},
and recently a detailed observation was made by the EGRET detector on
the Compton Gamma-Ray Observatory \cite{EGRET}. The EGRET observation
revealed that 
this cosmic GeV background is very hard with a photon energy index of
$\beta = 2.10 \pm 0.03$ ($dN/dE \propto E^{-\beta}$),
compared with
the background radiation in lower energy bands.

A large number of possible origins for the cosmic GeV background radiation
have been proposed (see Ref. \cite{EGRET} for a review). Currently the most
likely explanation is considered to be a superposition of unresolved 
active galactic nuclei of the blazar class. In fact, a large number of
gamma-ray emitting blazars have been observed by the EGRET, and the average
of the spectral index of these blazars is in good agreement with that of
the cosmic GeV background. 
However, there is no clear evidence that 
the GeV background is actually produced by unresolved blazars. In fact, 
majority of
radio selected blazars are likely to have luminosities falling off above
$\sim$ 10 GeV, and the observed GeV background 
extending to 100 GeV may be difficult to
explain by blazars\cite{bhattacharjee98}. 
Recently Mukherjee \& Chiang \cite{mukherjee} claimed that
blazars can explain only 25\% of the observed GeV background,
based on the luminosity function and its evolution of EGRET blazars,
and other sources of the GeV background must exist.
Therefore the origin of the cosmic GeV background radiation should still be
considered as an open question. 

In this paper we propose a new scenario
which can explain the extragalactic GeV background radiation by the 
intergalactic cascade of TeV gamma-rays emitted from cosmological gamma-ray
bursts (GRBs). Recently it has been pointed out that a very strong 
emission in TeV range is possible from GRBs by synchrotron radiation of
protons accelerated up to $10^{20}$ eV \cite{totani98a,totani98b}. There 
are also some recent observational suggestions for such strong emission of
TeV gamma-rays from GRBs \cite{Tibet,HEGRA}, and in fact they are well 
explained by the proton-synchrotron model \cite{totani98b}.

In section 2, we describe the proton synchrotron model of TeV gamma-ray
emission from GRBs. We then discuss, in section 3, the absorption of 
TeV gamma-rays
in the intergalactic field and show that GRBs can make a significant
contribution to the cosmic
GeV background. In section 4, we will discuss the spectrum of the 
GeV background radiation expected in this scenario and compare it to the
observation. Conclusions are presented in section 5.

\section{TeV Gamma-Ray Emission from Gamma-Ray Bursts}
Gamma-ray bursts (GRBs) are widely believed as
dissipation of kinetic energy of ultra-relativistic motion
produced by an expanding fireball
with a Lorentz factor of $\sim 10^2$--$10^3$
(see e.g., \cite{piran94} for a review). The recently discovered afterglows
following GRBs are also considered as similar phenomena,
which are dissipation
in the external shock generated by the collision with interstellar matter
\cite{afterglow}.
The radiation process of GRBs
is generally considered as electron synchrotron.
It has recently been pointed out that protons may carry a much larger
amount of energy than electrons in the shock-heated matter
by a factor of $\sim (m_p/m_e) = 2,000$, and most of this energy
is radiated as very high energy gamma-rays in TeV range by the synchrotron
radiation of protons accelerated to $\sim 10^{20}$ eV
\cite{totani98a}. Since the origin of the GRB energy is kinetic energy of
ultra-relativistic
motion, this situation is rather reasonable if the energy transfer from
protons into electrons is inefficient. 
The shock acceleration in the internal shocks of GRBs can 
accelerate protons up to $\sim 10^{20}$ eV. Synchrotron photons of such
ultra-high-energy protons are in 1--10 TeV range when the magnetic field
is in equipartition. The cooling time of such protons 
is sufficiently short ($\sim$ second), and hence
a significant fraction of energy carried by protons can be emitted as
TeV gamma-rays from GRBs, which would be much stronger emission than the
electron synchrotron in the ordinary photon energy range of 
GRBs (keV--MeV) (see ref. \cite{totani98a,totani98b} for detail).
In fact, although further 
observational confirmation is necessary, the Tibet\cite{Tibet}
and HEGRA \cite{HEGRA} groups have independently suggested the existence
of very strong emission above 10--20 TeV from some bright GRBs.
If these signals are truly from GRBs, they are
well explained by the above picture \cite{totani98b}.
Hence the TeV gamma-rays from GRBs may be the evidence for
the hypothesis that the ultra high energy cosmic rays (UHECRs)
observed on the Earth are produced by GRBs \cite{waxman95,usov95,vietri95}.

Since a typical amount of energy emitted from GRBs
in the ordinary soft gamma-ray range (keV--MeV) by electrons is
$\sim 10^{52-53} (\Delta \Omega / 4 \pi)$ ergs \cite{kulkarni98,totani98c}, 
where $\Delta \Omega$ is the opening solid anlge of GRB emission, 
a very large amount of energy [$\sim 10^{56}(\Delta \Omega / 4 \pi)$ erg]
is required for the engine of GRBs. 
Although it seems too large at first
glance, it is not theoretically impossible if GRB emission is strongly beamed.
The energy available by some magnetic processes such as the Blandford-Znajek
mechanism\cite{blandford77} is $\sim 10^{54}$ erg
from mergers of compact objects or collapses of massive stars
\cite{paczynski98,meszaros98}, which are currently considered as likely sources
of GRBs. If the GRB emission is strongly beamed with
$(\Delta \Omega / 4 \pi) \sim 10^{-2}$, the above energy can be explained.
There are also some very energetic models of GRBs producing 
$\sim 10^{56}$ ergs, by collapses of supermassive stars
\cite{prilutski75,fuller98}.
Therefore, in the following of this paper, we assume the above picture of GRBs,
i.e., 1) protons carry about a 2,000 times larger amount of
energy than electrons in the shocked region, and
2) protons are accelerated to $10^{20}$ eV and most of the energy 
carried by the ultra-high-energy protons is radiated in TeV range by
proton-synchrotron.

\section{Cascade of TeV Gamma-Rays and the Cosmic GeV Background}
Most of TeV gamma-rays emitted from GRBs would however be absorbed in 
the intergalactic space because of the $e^\pm$ creation with the cosmic
infrared background radiation\cite{stecker92}. The TeV gamma-rays observed
by the Tibet and HEGRA arrays are considered to be a very tiny fraction 
of the original flux from nearby GRBs ($z \ltilde 0.2$) \cite{totani98b}. 
On the other hand, the $e^\pm$ pairs created by the 
TeV photons in intergalactic fields, whose energy is also about TeV,
would scatter the 2.7K cosmic microwave background (CMB) photons
by the inverse-Compton process, and the energy of the secondary photons
is $\epsilon_{\gamma} = 0.6 \varepsilon_{\gamma, \rm TeV}^2$ GeV, where
$\varepsilon_{\gamma} = 10^{12} \varepsilon_{\gamma, \rm TeV}$ eV 
is the energy of primary photons. (For the simplicity, we neglect the 
redshift dependence of particle energies.) It should be noted that
created pairs would also scatter other low-energy background photons
such as the infrared band, 
but energy density of background radiation is dominated by
CMB and hence most of the energy of created pairs is lost in scattering
the CMB photons. 
It is uncertain whether we can  
directly observe these secondary photons
because the $e^\pm$ pairs would be bent by intergalactic magnetic fields.
The created pairs run about $l = 0.35
\varepsilon_{\rm e, TeV}^{-1}$ Mpc before they cool down by the inverse-Compton
scattering of CMB photons, 
where $\varepsilon_{\rm e} = 10^{12} \varepsilon_{\rm e, TeV}$
eV is the energy of created pairs. The Larmor radius of pairs is given by
$r_L = 1.1 \varepsilon_{\rm e, TeV} B_{-12}^{-1}$ kpc, 
where $B = 10^{-12} B_{-12}$ G is the
intergalactic magnetic field. In order to observe the inverse-Compton
photons within a time delay of about day, a very small bending angle of
$(l/r_L) \alt 10^{-6} d_3^{-1/2}$ rad, i.e., $B \alt 10^{-20}$ G
is required, where $d = 3000 d_3$ Mpc is the distance to the source. 
This value is much smaller than the current upper limits on the intergalactic
magnetic fields, $B_{\rm IGM} \ltilde 10^{-9} (L_{\rm rev}/ {\rm Mpc})^{-1/2}
\rm \ G$, where $L_{\rm rev}$ is the field reversal scale \cite{kronberg94}. 
At least a magnetic field of order $\sim 10^{-20}$
G is expected in intergalactic fields during the structure formation in the
universe by thermoelectric currents \cite{kulsrud96}. If the intergalactic
magnetic field is still at this level at the present time, 
the secondary GeV gamma-rays may be marginally observable. On the other hand,
it should also be noted that a magnetic field of
$\sim 10^{-12}$ G is necessary in the intergalactic field
for the hypothesis that GRBs are the origin
of UHECRs, in order to make the arrival time of UHECRs on the Earth
sufficiently dispersed (i.e., not like a burst)\cite{waxman95}.
In our model, protons are accelerated to $10^{20}$ eV in order to 
emit TeV gamma-rays from GRBs, and if these ultra-high-energy protons 
also explain UHECRs, it seems difficult to observe
the secondary GeV photons.

In any case, such secondary GeV photons should form a uniform cosmic
background radiation at the present time. 
In the following we show that the flux
and spectrum predicted by our scenario are consistent with
the recent observations of the cosmic GeV background by the EGRET 
experiment\cite{EGRET}. 
First we give an order-of-magnitude estimation for the GeV background flux.
The observed energy density of GeV background photons is 
$\sim 6 \times 10^{-18} \ \rm
erg \ cm^{-3}$ in 30 MeV--30 GeV\cite{EGRET}. In our scenario this energy
density is a sum of the energy emitted from all GRBs which have ever occurred
in the universe.
The occurrence rate of GRBs depends on the unknown distance scale of
GRBs, and it is about $\sim 10^{-9} b \ \rm yr^{-1}
Mpc^{-3}$ \cite{totani98c,totani97} with typical distance 
scales of cosmological 
GRBs ($z_{\max} \sim$ 3), where 
$b = 4 \pi / \Delta \Omega$ is the beaming factor and 
$z_{\max}$ is the redshift of the most
distant GRBs observed by the BATSE experiment\cite{BATSE}.
By using this GRB occurrence rate and assuming the age of the
universe as 15 Gyr, the total number of GRBs which have ever occurred is
$\sim 15 b \ \rm Mpc^{-3}$. 
(The evolutionary effect of GRB rate may somewhat change this estimate, 
but here we neglect it in this order-of-magnitude estimate.)
Hence the total energy of GeV gamma-rays which should be produced by
one GRB event becomes $\sim 10^{55} b^{-1}$ erg. Since we are now
considering the case that the total GRB energy is $\sim 10^{56}b^{-1}$ erg, 
this energy can be supplied from GRBs. 
In fact, this energy is about $10^3$ times larger
than the conventional energy estimate of GRBs, $\sim 10^{52} b^{-1}$ erg. 
Therefore
it has been considered that GRBs cannot be the origin of the extragalactic
GeV background. However, in the present model, protons carry $(m_p/m_e)
\sim 2,000$ times larger amount of 
energy than electrons, and significant fraction of this energy
will be emitted as TeV gamma-rays and eventually converted into GeV photons.
Hence the shortage of energy is
just compensated by the proton-electron mass ratio, 
and now the energy production
of GRBs is in agreement with that of the cosmic GeV background.

It is also interesting to compare the energy production rate estimated above
with that of UHECRs observed on the earth.
The energy production rate of UHECRs per one GRB
is about $\sim 10^{54} b^{-1}$ erg
(see eq. 12 of Ref. \cite{totani98b}), and hence the energy production rate
of the GeV background is somewhat
larger than that of UHECRs. This suggests that 
the photons produced by the intergalactic cascade originated by UHECRs
cannot be a dominant component of the GeV background. 
Instead, ultra high energy protons lose their energy mainly by
synchrotron radiation in GRBs, and most of their energy is emitted as
TeV gamma-rays. The dominant component of the GeV background is the 
secondary GeV photons produced by synchrotron TeV photons from GRBs,
and only a small fraction of $10^{20}$ eV protons can escape 
from GRBs to be observed as UHECRs.

\section{Spectrum of the Cosmic GeV Background}

Next we consider the spectrum of the GeV background predicted by this model.
It is not straightforward to calculate the spectrum of the GeV
background generated by intergalactic cascade of higher energy photons.
The intergalactic optical depth of high energy gamma-rays changes
significantly with the photon energy, depending on the flux, spectrum,
and thier evolution of the cosmic infrared background, which are
still uncertain. 
It is known that
the final spectrum of the gamma-ray background is generally
insensitive to the original spectrum of gamma-rays emitted from the source
\cite{coppi97}. 
Coppi and Aharonian \cite{coppi97} have made a realistic calculation
of the spectrum of the GeV background produced by the cascade of TeV
photons, taking into account various reaction processes in intergalactic
fields and realistic properties of the cosmic infrared background. 
According to their calculation, 
most of the cascade energy produced by very high energy gamma-rays
above $\sim$ TeV is contained in the background radiation in the EGRET range.
The spectrum of the cascade photons becomes about $\beta \sim$ 1.8--2
above the energy $\epsilon_b \sim [\epsilon_{\rm cut} (z_s) / {\rm 1 TeV}]^2 
\rm \ GeV$, where $\epsilon_{\rm cut}$ is the cut off energy at which 
an observed spectrum
of very high energy gamma-rays from a source at $z = z_s$ shows a cut-off
due to the pair creation process. At $\epsilon_\gamma < \epsilon_b$, 
the spectral index becomes
$\beta \sim 1.5$. Based on a recent realistic  calculation
\cite{salamon98}, $\epsilon_{\rm cut} \sim$ 500, 50, and 20 GeV 
for $z_s$ = 0.1, 1, and 3, respectively. Therefore for a typical cosmological
GRBs with $z \gtilde 1$, $\epsilon_b$ is well below 30 MeV at which the hard 
component of the cosmic GeV background starts. Therefore the spectrum
of the cascade photons becomes $\beta \sim 2$ in the EGRET range and it
is fully consistent with the EGRET observation.

The cosmological evolution
of GRB rate may also affect the spectrum of the GeV background through
the cosmological redshift effect. More detailed calculation including
a realistic rate evolution of GRBs and various reaction processes
in the intergalactic field is necessary
to verify whether the GeV spectrum predicted by our scenario is 
consistent with the observation, and this point will be studied in
future publications. 

\section{Conclusions}
We have shown that the extragalactic background radiation in the GeV range
can be explained by GRBs, by the intergalactic cascade of TeV photons 
from GRBs emitted by synchrotron radiation of $10^{20}$ eV protons.
For this scenario, protons must carry about $(m_p/m_e)$ times larger 
amount of energy 
than electrons, and it is a reasonable situation if the energy 
transfer from protons into electrons in shocked region is inefficient.
In order to distinguish this scenario from other scenarios for the cosmic
GeV background, more precise estimate of the background spectrum is necessary
both in the theoretical modeling and in observation.
As we have mentioned earlier in this paper, majority of
radio selected blazars are likely to have luminosities falling off above
$\sim$ 10 GeV, and the observed GeV background 
extending to 100 GeV may be difficult to
explain by blazars\cite{bhattacharjee98}. 
On the other hand, in our scenario the GeV gamma-ray background is
a consequence of the cascade of much higher energy photons, and
hence extension of the spectrum up to $\sim$ 100 GeV is naturally expected.
This point may be a crucial point to discriminate between the two scenarios.

The author has been financially supported by the JSPS fellowships 
and scientific research fund of Ministry of Education, Science, 
and Culture in Japan.



\begin{references}
\bibitem{SAS}
C.E. Fichtel, et al. Astrophys. J, 217, L9 (1977);

\bibitem{EGRET} 
P. Sreekumar, et al. Astrophys. J. { 494}, 523 (1998) 

\bibitem{bhattacharjee98}
P. Bhattacharjee, Q. Shafi, and F.W. Stecker, 
Phys. Rev. Lett. { 80}, 3698 (1998).

\bibitem{mukherjee}
R. Mukherjee and J. Chiang, Astroparticle Phys. in press
(astro-ph/9902003).

\bibitem{totani98a}
T. Totani, Astrophys. J. { 502}, L13 (1998). 

\bibitem{totani98b}
T. Totani, to appear in Astrophys. J. Lett., 
astro-ph/9810206

\bibitem{Tibet}
M. Amenomori, et al., Astron. Astrophys. { 311}, 919 (1996)

\bibitem{HEGRA}
L. Padilla et al., Astron. Astrophys. { 337}, 43 (1998)

\bibitem{piran94} T. Piran, in Unsolved Problems in Astrophysics,
ed. J.N. Bahcall and J.P. Ostriker (New Jersey: Princeton University Press),
343 (1994)

\bibitem{afterglow} B. Paczy\'nski, and J. Rhoads, Astrophys. J., 
418, L5 (1993); J.I. Katz, Astrophys. J., 432, L107 (1994);
P. M\'esz\'aros and  M.J. Rees, M., Astrophys. J., 476, 232 (1997);
M. Vietri, Astrophys. J., 478, L9 (1997)

\bibitem{waxman95}
E. Waxman,
Phys. Rev. Lett. { 75}, 386 (1995). 

\bibitem{usov95}
M. Milgrom and V. Usov, Astrophys. J. 449, L37 (1995)

\bibitem{vietri95}
M. Vietri, 
Astrophys. J. { 453}, 883 (1995).

\bibitem{kulkarni98}
S.R. Kulkarni, et al., Nature { 393}, 35 (1998).

\bibitem{totani98c}
T. Totani,
Astrophys. J. in press. astro-ph/9805263 (1998).

\bibitem{blandford77}
R.D. Blandford and R.L. Znajek, Mon. Not. Roy. Astron. Soc. { 179}, 
433 (1977).

\bibitem{paczynski98}
B. Paczy\'{n}ski, Astrophys. J. { 494}, L45 (1998).

\bibitem{meszaros98}
P. M\'esz\'aros, M.J. Rees, and R.A.M.J. Wijers, preprint, (astro-ph/9808106)

\bibitem{prilutski75}
O.F. Prilutski and V.V. Usov, Astrophys. and Sp. Science 34, 395 (1975)

\bibitem{fuller98}
G.M. Fuller and X. Shi, ApJL, 502, L5 (1998)

\bibitem{stecker92}
F.W. Stecker, O.C. deŽ Jager,  and M.H. Salamon,  
Astrophys. J. { 390}, L49 (1992), and
references therein.  

\bibitem{kronberg94}
P.P. Kronberg, Rep. Prog. Phys. 57, 325 (1994)

\bibitem{kulsrud96}
R.M. Kulsrud, R. Cen, J.P. Ostriker, and D. Ryu, Astrophys. J. 480, 481 (1996)

\bibitem{totani97}
T. Totani, Astrophys. J. { 486}, L71 (1997).

\bibitem{BATSE}
C. Meegan et al., Astrophys. J. Suppl. 106, 65 (1996)

\bibitem{salamon98}
M.H. Salamon and F.W. Stecker, Astrophys. J. 493, 547 (1998)

\bibitem{coppi97}
P. S. Coppi and F. A. Aharonian, Astrophys. J. 487, L9 (1997)

\end{references}
\end{document}